\documentclass[conference]{IEEEtran}
\IEEEoverridecommandlockouts
\usepackage{cite}
\usepackage{amsmath,amssymb,amsfonts}
\usepackage{graphicx}
\usepackage{textcomp}
\usepackage{xcolor}
\usepackage{CJKutf8}
\usepackage{colortbl}
\usepackage{makecell}
\usepackage{hyperref}
\usepackage{balance}
\usepackage{ulem}
\hypersetup{
    colorlinks=true,
    linkcolor=blue,}
\usepackage{booktabs}
\usepackage{multirow}
\usepackage{algorithm}  
\usepackage{amssymb}
\usepackage{algpseudocode}  
\usepackage[misc]{ifsym}
\usepackage{amsthm}
\usepackage{epstopdf}
\usepackage{pifont}

\newcolumntype{P}[1]{>{\centering\arraybackslash}p{#1}}
\def\BibTeX{{\rm B\kern-.05em{\sc i\kern-.025em b}\kern-.08em
    T\kern-.1667em\lower.7ex\hbox{E}\kern-.125emX}}
\begin{document}

\title{HiNet: Novel Multi-Scenario \& Multi-Task Learning with Hierarchical Information Extraction
}

\author{

\IEEEauthorblockN{
Jie Zhou\IEEEauthorrefmark{1}\IEEEauthorrefmark{2}\textsuperscript{+}, 
Wenhao Li\IEEEauthorrefmark{2}\textsuperscript{+},
Xianshuai Cao\IEEEauthorrefmark{2},
Lin Bo\IEEEauthorrefmark{2},
Kun Zhang\IEEEauthorrefmark{2}, 
Chuan Luo\IEEEauthorrefmark{1}\textsuperscript{\Letter}\thanks{\textsuperscript{+} Equal contribution.}\thanks{\textsuperscript{\Letter} Corresponding authors.}, 
Qian Yu\IEEEauthorrefmark{1}\textsuperscript{\Letter}
}
\IEEEauthorblockA{\IEEEauthorrefmark{1}School of Software, Beihang University, Beijing, China}
\IEEEauthorblockA{\IEEEauthorrefmark{2}Meituan, Beijing, China}
\IEEEauthorblockA{\{zhoujiee, chuanluo, qianyu\}@buaa.edu.cn, \{liwenhao22, caoxianshuai, bolin02, zhangkun32\}@meituan.com} 

}

\maketitle

\begin{abstract}
Multi-scenario \& multi-task learning has been widely applied to many recommendation systems in industrial applications, wherein
an effective and practical approach is to carry out multi-scenario transfer learning on the basis of the Mixture-of-Expert (MoE) architecture. 
However, the MoE-based method, which aims to project all information in the same feature space, cannot effectively deal with the complex relationships inherent among various scenarios and tasks, resulting in unsatisfactory performance. 
To tackle the problem, we propose a \uline{H}ierarchical \uline{i}nformation extraction \uline{Net}work (HiNet) for multi-scenario and multi-task recommendation, which achieves hierarchical extraction based on coarse-to-fine knowledge transfer scheme.
The multiple extraction layers of the hierarchical network enable the model to enhance the capability of transferring valuable information across scenarios while preserving specific features of scenarios and tasks. 
Furthermore, a novel scenario-aware attentive network module is proposed to model correlations between scenarios explicitly.
Comprehensive experiments conducted on real-world industrial datasets from Meituan Meishi platform demonstrate that HiNet achieves a new state-of-the-art performance and significantly outperforms existing solutions.
HiNet is currently fully deployed in two scenarios and has achieved 2.87\% and 1.75\% order quantity gain respectively.
\end{abstract}
\normalem
\section{Introduction}

\begin{figure}[t]
  \centering
  \includegraphics[width=0.95\linewidth]{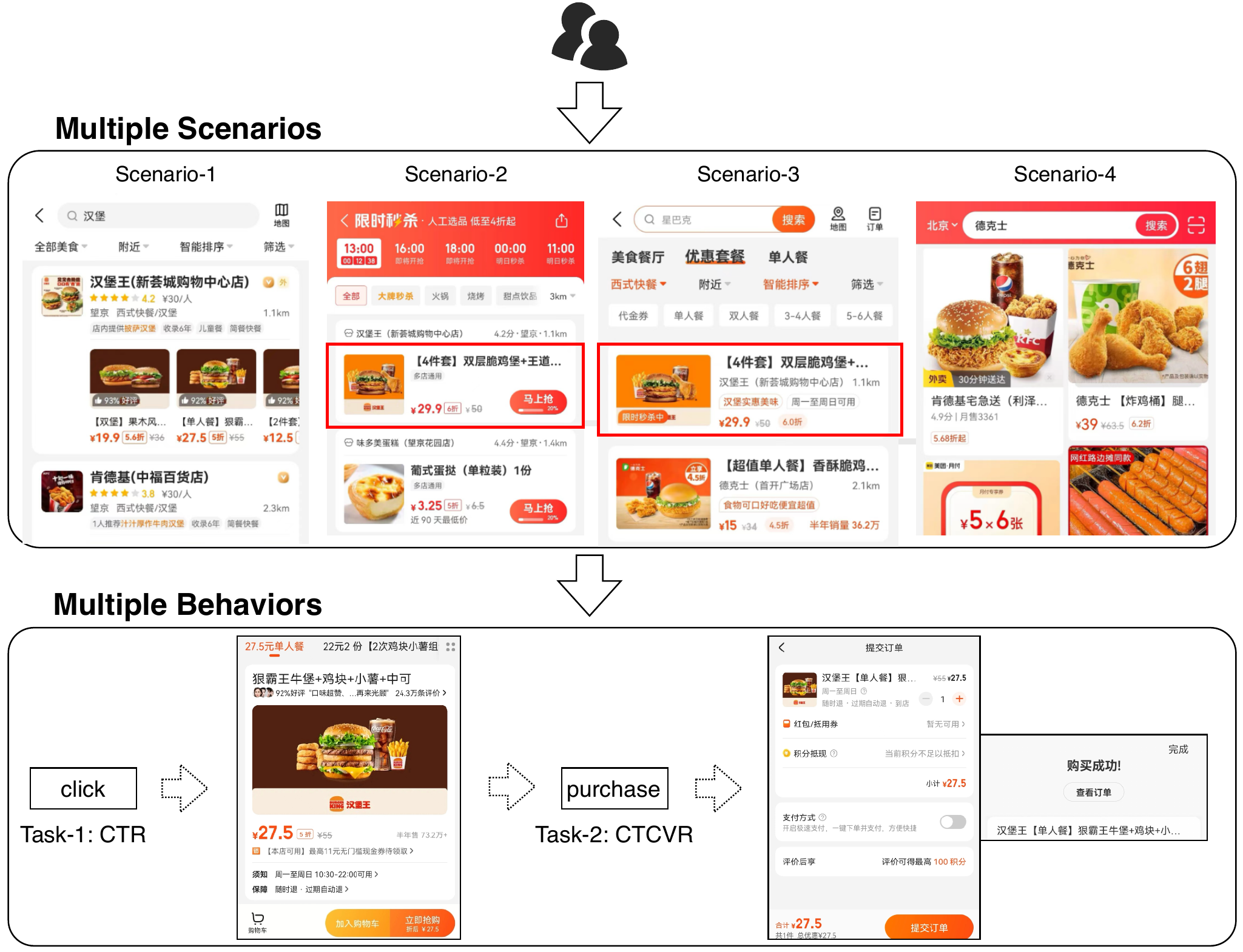}
   \caption{
   Illustration of user interaction flow inside the Meituan Meishi application. Users may access multiple scenarios listed in the figure. As shown at the bottom, users enter the item details page via the click behavior and then complete the item purchase through the order submission behavior, corresponding to the \textit{CTR} and \textit{CTCVR} task in the recommendation system. In addition, the red rectangles represent that the same item may be presented to the user in multiple scenarios.}
  \label{fig:user_behave}
\end{figure}

With the explosive growth of online information and services, recommendation systems have been a pivotal tool for enterprises to provide personalized suggestions for their customers in order to overcome information overload and improve the quality of decision-making process.
Such recommendation systems usually are developed respectively under each certain scenario, due to scenario-specific data distribution and feature space.
However, in real-world industrial platforms, there are a variety of scenarios (e.g., homepage feeds, banner, etc.) that items are ranked and presented to the user based on personalized recommendation models.
E.g., in the Meituan Meishi application, the users may have multiple behavior interactions across multiple scenarios on the platform. 
As the number of recommendation scenarios increases, the traditional approach that trains a single ranking model for each scenario only based on data of itself, can lead to the following issues: 
1) Modeling scenario solely on its own data can not utilize shared knowledge across scenarios, especially considering that there exists overlapping users and items in multiple scenarios. 
2) Long-tail scenarios with low traffic and sparse user feedback do not have a sufficient amount of data to train a scenario-specific ranking model efficiently.
3) The ranking model for each scenario needs to be trained and deployed independently, which heavily increases the computational cost and maintenance burden. 
While modeling multiple scenarios separately may cause the above issues, simply training one shared ranking model on merged data can not effectively capture the distinctive properties of each scenario.

It should be noted that, besides the multi-scenario recommendation problem, user satisfaction and engagement in each scenario usually have diverse metrics that needs to be optimized simultaneously, e.g. click-through rate (CTR) and click-through \& conversion rate (CTCVR). Therefore, it is crucial to develop an effective and unified framework to resolve the complication of optimizing various metrics in multiple scenarios.

Recent advance \cite{li2020improving,sheng2021one} in addressing this problem is to model the multi-scenario recommendation as a multi-task learning (MTL) problem and mainly focus on investigating the multi-gate mixture-of-experts (MMoE)\cite{ma2018modeling} architecture to learn commonalities and distinctions between scenarios.
However, the MTL-based methods that project data from multiple scenarios into the same feature space, can not sufficiently capture the complex relations among scenarios with multiple tasks.

In this paper, we propose a novel \uline{H}ierarchical \uline{i}nformation extraction \uline{Net}work (HiNet) for multi-scenario \& multi-task recommendation problem, based on the intuition that scenario-related and task-related information are at different levels of granularity thus should by processed hierarchically.
Specifically, we design an end-to-end two-layer framework that jointly models inter-scenario and inter-task information sharing and collaboration. 
First, in the scenario extraction layer, HiNet is able to extract scenario-shared and scenario-specific information through separate expert modules. 
To further enhance the representation learning of current scenario, the scenario-aware attentive network is designed to explicitly learn contributions from other scenarios to the current one. 
Then, in the task extraction layer, we utilize a customized gate control network consisting of task-shared and task-specific expert networks, which effectively alleviates parameter interference between shared and task-specific knowledge in the multi-task learning\cite{tang2020progressive}.
By separating scenario-level and task-level information extraction at the model structure, multiple tasks in various scenarios can be explicitly divided into different feature spaces for optimization, which will facilitate the improvement of model performance.

The main contributions of this paper are as follows:
\begin{itemize}
    \item [1)] We propose HiNet, a novel multi-scenario \& multi-task learning model, for optimizing various metrics in multiple recommendation scenarios, which innovatively applies hierarchical information extraction structure. 
    \item [2)] On the basis of hierarchical information extraction structure, we propose the scenario-aware attentive network (SAN) module to enhance the ability of modeling complex correlations between scenarios explicitly.
    \item [3)] Empirical results from offline evaluation and online A/B tests prove the superiority of HiNet over state-of-the-art models. Currently, HiNet has been fully deployed in two scenarios at Meituan.
\end{itemize}
\section{Related Work}

\begin{figure*}[t]
  \centering
  \includegraphics[width=0.8\linewidth]{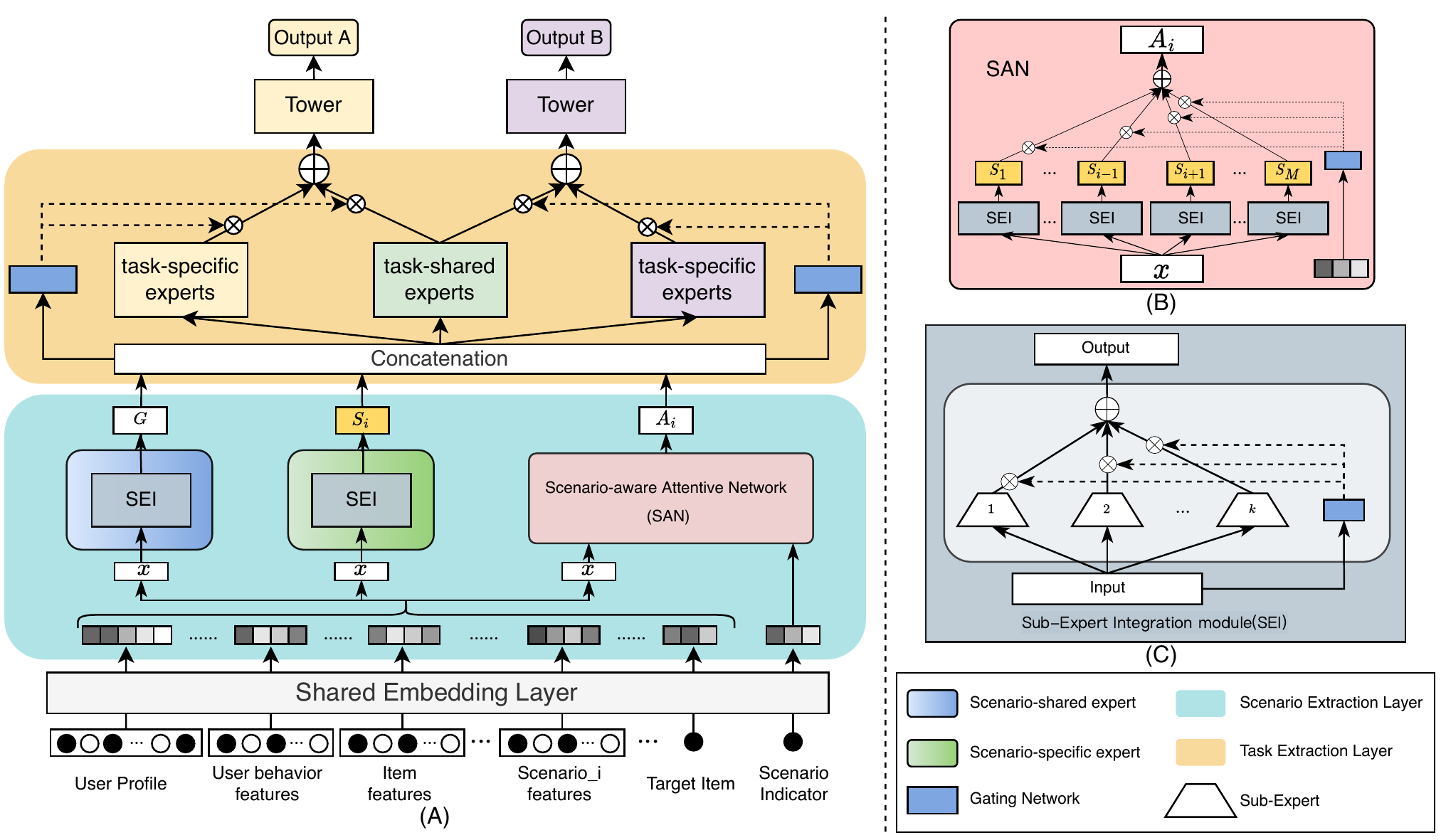}
  \caption{
  (A) The architecture of the proposed HiNet, which utilizes Scenario Extraction Layer and Task Extraction Layer to obtain scenario and task representations, respectively. (B) At the scenario extraction layer, the Scenario-aware Attentive Network (SAN) module is designed to enhance the representation learning process of current scenario in addition to the scenario sharing/specific experts. The CGC module is adopted at the task extraction layer to further extract task information. (C) The SEI (Sub-Expert Integration) module is used in the scenario extraction layer. 
  }
  \label{fig:architecture}
\end{figure*}

This work is mainly inspired by advanced multi-scenario and multi-task learning research. The following subsections mainly introduce the contents related to our research work.

\subsection{Multi-Scenario Learning}
Multi-scenario learning is mainly designed to solve the learning problem in different scenarios, which models the shared and specific information among scenarios with the help of transfer learning \cite{torrey2010transfer,pan2010survey}. 
Since there are certain similarities between multi-scenario and multi-task learning, Li \emph{et al.} proposed HMoE \cite{li2020improving}, where HMoE takes advantage of Multi-task Mixture-of-Experts to implicitly identify distinctions and commonalities between tasks in the feature space, and improves the performance with a stacked model learning task. Structurally, HMoE projects the data in the multi-scenario formulation into the same feature space, and then use the method of multi-task learning to optimize. However, the actual multi-scenario datasets are usually complex, resulting in the poor ability of the above methods to capture the shared and specific information of each scenario.
To model the shared and specific features among scenarios explicitly, Sheng \emph{et al.}\cite{sheng2021one} proposes the star topology (STAR), which contains a shared network and a scenario-specific network structure designed for each scenario, and the goals of each scenario are learned through the shared network and the scenario-specific expert network. 
STAR trains a single model to serve all scenarios by leveraging data from all scenarios simultaneously, capturing the characteristics of each scenario, and modeling the commonalities among different scenarios. According to \cite{zhu2021cross}, due to the correlation among scenarios, each of other scenario has a certain representation ability for the current scenario. However, STAR does not take this into account, which may lack the information representations of scenarios. 

\subsection{Multi-task Learning}

Unlike multi-scenario learning, multi-task learning \cite{zhou2023feature,ruder2017overview,zhou2023dcrnn,rosenbaum2018routing,wang2013online} focuses on the ability to model multiple tasks in a single scenario. Shared Bottom \cite{caruana1997multitask} is a simple multi-task learning model that concentrates on sharing the underlying feature information to model different tasks. However, Shared Bottom is a hard parameter sharing model, which limits the learning capability of the model for different tasks. 
To further improve the modeling capability for different tasks, Ma \emph{et al.} \cite{ma2018modeling} proposed MMoE. MMoE expresses the underlying feature information in multiple expert networks, follows the idea of ensemble learning to improve the representation capability of the model. MMoE uses a flexible gating network to construct information representations for different tasks, and then pass the representation results to the \textit{tower unit} for the corresponding task prediction. 
However, as the model continues to iterate, MMoE will experience the \textit{seesaw} phenomenon, which is considered as a manifestation of the negative transfer problem \cite{weiss2016survey,torrey2010transfer,pan2010survey}. Therefore, Tang \emph{et al.} \cite{tang2020progressive} proposed Progressive Layered Extraction (PLE) to address the problem. PLE separates shared components and task-specific components explicitly and adopts a progressive routing mechanism to extract and separate deeper semantic knowledge gradually, improving efficiency of joint representation learning and information routing across tasks in a general setup. 

In general, the multi-scenario model can only handle the single-task learning problem in multiple scenarios while the multi-task model is only able to address the negative transfer among multiple tasks in a single scenario. 
In this paper, we combine the multi-scenario and the multi-task model to transfer valuable information among multiple scenarios while effectively dealing with the negative transfer problem across multiple tasks.

\section{The HiNet Approach}
In this section, we introduce the technical details of our proposed HiNet approach.

\subsection{Problem Formulation}
As discussed above, we focus on the essential multi-scenario \& multi-task recommendation problem for platforms serving multiple scenarios. 
We define the problem as: $\hat{y}_i^j=f_i^j\left(x, s_i\right)$, where $s_i$ denotes the $i$-th scenario indicator, $\hat{y}_i^j$ is the prediction of task-$j$ under the $i$-th scenario and $x$ denotes the input dense features. Note that the original input features include: user profile, user behavior features, current scenario specific features and item features, of which the numerical features are first transformed into categorical features. Then $x$ can be obtained by mapping all categorical features to a low-dimensional vector space. 
As for the specific optimization target, CTR and CTCVR tasks are considered for each scenario. 

\subsection{Hierarchical Information Extraction Network}
The architecture of the proposed HiNet model (as shown in Fig.~\ref{fig:architecture}) is composed of 2 layers: Scenario Extraction Layer and Task Extraction Layer. 

\subsubsection{Scenario Extraction Layer}
This layer plays the role of transferring and sharing valuable information among scenarios as well as extracting scenario-specific characteristics, which is the basis for improving performance at higher layer.
There includes scenario-shared expert network, scenario-specific expert network, and scenario-aware attentive network.

\paragraph{Scenario-shared/specific Expert Network}

Considering overlapping user interactions and items across scenarios, there exists shared information among data of multiple scenarios. Strategically, we design scenario-shared expert network. 
Note that inspired by MoE \cite{shazeer2017outrageously,2013Learning,jordan1994hierarchical,jacobs1991adaptive,yuksel2012twenty}, the scenario-shared expert is generated by using the Sub-Expert Integration (SEI, as shown SEI module in Fig.~\ref{fig:architecture}) module. Specifically, the output $G$ of scenario-shared expert is formulated as: 
\begin{equation}
    G = \sum_{k=1} ^{K_{s}} g_{sh}^{k}(x) Q^k_{sh}(x)
\end{equation}
where $Q^k_{sh}$ denotes the $k$-th sub-expert network composed of multi-layer perceptron (MLP) with activation function, $K_{s}$ is the number of sub-experts $Q_{sh}(\cdot)$, and $g_{sh}^{k}(x)$ denotes the output of gating network, which is a simply linear transformation of $x$ with a softmax layer:
\begin{equation}
    g_{sh}^{k}(x) = softmax(W_{sh}^{k}x)
\end{equation}

In addition to the extraction of shared information, we also design a scenario-specific expert network (composed of SEI module) for each scenario to learn scenario-specific information. 
The output $S_{i}$ of scenario-specific expert network of $i$-th scenario is represented as follows:
\begin{equation}
    S_{i} = \sum_{k=1} ^{K_{i}} g_{sp}^{k}(x) Q_{sp}^{k}(x)
\end{equation}
where $Q_{sp}^{k}$ denotes the $k$-th scenario-specific sub-expert network of the $i$-th scenario, $K_{i}$ is the number of $Q_{sp}(\cdot)$, and $g_{sp}(x)$ denotes output of scenario-specific gating network.

\paragraph{Scenario-aware Attentive Network}
As mentioned above, there exists correlations among scenarios, thus the information from other scenarios can also contribute to the current one, which complements the current scenario's representation\cite{zhang2022diverse}.
Consider that scenarios have varying contributions to each other, we design the Scenario-aware Attentive Network (SAN) module to measure the importance of the information from other scenarios to the current one. 
Specifically, SAN contains two parts of input: the embedding vector $Emb(s_i)$ of the scenario indicator $s_i$, which is used to calculate the importance weights of other scenarios through a gating network with a softmax function,
and $\textbf{S}=\left[S_1,\cdots,S_{i-1},S_{i+1},\cdots, S_M\right]$ corresponding to representation of scenarios obtained by a series of SEI modules. 
The output $A_i$ of $i$-th scenario through SAN is the weighted sum of scenarios' representations $\textbf{S}$:
\begin{equation}
    A_i = \sum_{m\ne{i}}^{M} g_{a}^{i}(Emb(s_i))  S_{m}
\end{equation}
\begin{equation}
    g_{a}^{i}(x) = softmax(W_{a}^{i} Emb(s_i))
\end{equation}
where $Emb(\cdot)$ denotes that the scenario indicator is projected as the embedding vector, $g_{a}^{i}(\cdot)$ represents the gating network based on weight $W_{a}^{i} \in R^{(M-1) \times d} $, $d$ denotes the dimension of the $Emb(\cdot)$, and $M$ is the number of scenarios.

In fact, the SAN module can transfer information across scenarios to varying degrees depending on complex scenario correlations, which effectively enhances representation learning of scenarios and improves the performance of HiNet.

To sum up, the overall output $C_{i}$ of the scenario extraction layer can be expressed as: $C_{i} = Concat\left[ G, S_i, A_i\right]$.

\begin{table}[ht]
  \caption{Statistics of the dataset for all scenarios.}
  \centering
  \label{tab:tb1}
  \resizebox{1.0\columnwidth}{!}{
  \begin{tabular}{c|ccccccc}
    \hline & All  & Scenario-$a$ & Scenario-$b$ & Scenario-$c$ & Scenario-$d$ & Scenario-$e$ & Scenario-$f$ \\
    \hline 
    \#Exposure & 23.6M & 11.5M & 6.3M & 1.5M & 1.7M & 89K & 1.7M \\
    \#Click & 3.5M & 1.4M & 1.4M & 204K & 234K & 3.6K & 188K \\
    \#Order & 715K & 302K & 375K & 9.3K & 9.1K & 0.28K & 19K \\
    \hline
    \#CTR & 15.38\% & 12.56\% & 22.50\% & 14\% & 13.84\% & 4.12\% & 11.17\% \\
    \#CTCVR & 3.15\% & 2.64\% & 5.91\% & 0.64\% & 0.54\% & 0.31\% & 1.13\% \\
    \hline
  \end{tabular}
  }
\end{table}

\begin{table*}[ht]
  \caption{Performance comparison of the methods in all scenarios, the difference between HiNet and the baseline was statistically significant at the 0.05 level.}
  \centering
  \label{tab:tb2}
  \resizebox{0.9\linewidth}{!}{
  \begin{tabular}{c|c|cccccccccccc}
  \hline \multirow{2}{*} { Model Type } & \multirow{2}{*} { Model } & \multicolumn{2}{c} { Scenario-$a$ } & \multicolumn{2}{c} { Scenario-$b$ } & \multicolumn{2}{c} { Scenario-$c$ }& \multicolumn{2}{c} { Scenario-$d$ }& \multicolumn{2}{c} { Scenario-$e$ }& \multicolumn{2}{c} { Scenario-$f$ } \\
		\cline { 3 - 14 } ~ & ~ & auc\_ctr & auc\_ctcvr & auc\_ctr & auc\_ctcvr & auc\_ctr & auc\_ctcvr & auc\_ctr & auc\_ctcvr & auc\_ctr & auc\_ctcvr & auc\_ctr & auc\_ctcvr \\
    \hline 
    \multirow{3}{*} { Multi-task learning } & Shared Bottom & 0.7631 & 0.8426 & 0.7863 & 0.7883 &  0.6962 & 0.7307 & 0.6192 & 0.6131 & 0.6108 & 0.6079 & 0.7847 & 0.8088 \\
    ~ & MMoE & 0.7876 & 0.8581 & 0.7955 & 0.7978 & 0.7098 & 0.7532 & 0.6702 & 0.6881 & 0.6203 & 0.6106 & 0.8096 & 0.8337 \\
    ~ & PLE & 0.7741 & 0.8600 & 0.7936 & 0.7852 & 0.7108 & 0.7516 & 0.6794 & 0.7053 & 0.6169 & 0.6108 &  0.8075 & 0.8295 \\
    \hline
    \multirow{2}{*} { Multi-scenario learning } & HMoE & 0.7873 & 0.8556 & 0.7947 & 0.8014 & 0.7015 & 0.7531 & 0.6754 & 0.7174 & 0.6088 & 0.6183 & 0.8073 & 0.8311 \\
    ~ & STAR & 0.7716 & 0.8572 & 0.7906 & 0.8138 & 0.7040 &  0.7499 & 0.6786 & 0.7144 & 0.6163 & 0.6107 & 0.8051 & 0.8305 \\
    \hline
    ours & HiNet & \textbf{0.7904} & \textbf{0.8675} & \textbf{0.7996} & \textbf{0.8194} & \textbf{0.7146} & \textbf{0.7620} & \textbf{0.6858} & \textbf{0.7282} & \textbf{0.6253} & \textbf{0.6203} & \textbf{0.8115} & \textbf{0.8359} \\
    \hline
  \end{tabular}
  }
\end{table*}
\begin{table*}[ht]
  \caption{Ablation study of the HiNet model.}
  \centering
  \label{tab:tb4}
  \resizebox{0.9\linewidth}{!}{
  \begin{tabular}{c|cccccccccccc}
    \hline \multirow{2}{*} { Model } & \multicolumn{2}{c} { Scenario-$a$ } & \multicolumn{2}{c} { Scenario-$b$ } & \multicolumn{2}{c} { Scenario-$c$ }& \multicolumn{2}{c} { Scenario-$d$ }& \multicolumn{2}{c} { Scenario-$e$ }& \multicolumn{2}{c} { Scenario-$f$ } \\
			\cline { 2 - 13 } & auc\_ctr & auc\_ctcvr & auc\_ctr & auc\_ctcvr & auc\_ctr & auc\_ctcvr & auc\_ctr & auc\_ctcvr & auc\_ctr & auc\_ctcvr & auc\_ctr & auc\_ctcvr \\
    \hline 
    HiNet(w/o hierarchy) & 0.7737 & 0.8574 & 0.7836 & 0.8028 & 0.7108 & 0.7503 & 0.6753 & 0.6981 & 0.6151 & 0.6122 & 0.7861 & 0.8170 \\
    HiNet(w/o SAN) & 0.7880 & 0.8579 & 0.7947 & 0.8148 & 0.7114 & 0.7545 & 0.6817 & 0.7252 & 0.6197 & 0.6198 & 0.8045 & 0.8287 \\
    HiNet(w/o task gating) & 0.7756 & 0.8555 & 0.7754 & 0.7955 & 0.6822 & 0.7141 & 0.6518 & 0.6904 & 0.6162 & 0.6074 & 0.7827 & 0.8178 \\
    HiNet(w/o scenario gating) & 0.7792 & 0.8591 & 0.7763 & 0.7980 & 0.6968 & 0.7416 & 0.6535 & 0.7075 & 0.6177 & 0.6101 & 0.7946 & 0.8195 \\
    HiNet(w/o scenario \& task gating) & 0.7716 & 0.8528 & 0.7722 & 0.7919 & 0.6581 & 0.6842 & 0.6435 & 0.6838 & 0.6008 & 0.6007 & 0.7544 & 0.7786 \\
    \hline
    HiNet(ours) & \textbf{0.7904} & \textbf{0.8675} & \textbf{0.7996} & \textbf{0.8194} & \textbf{0.7146} & \textbf{0.7620} & \textbf{0.6858} & \textbf{0.7282} & \textbf{0.6253} & \textbf{0.6203} & \textbf{0.8115} & \textbf{0.8359} \\
    \hline
  \end{tabular}
  }
\end{table*}

\subsubsection{Task Extraction Layer}
To address the negative transfer problem in multi-task learning, in task extraction layer, we employ the Customized Gate Control (CGC) inspired by \cite{tang2020progressive}. 
\paragraph{Customized Gate Control} 
This module is mainly composed of two parts: task-shared expert networks and task-specific expert networks. 
The former is primarily responsible for learning shared information in all tasks of the current scenario, while the latter is used to extract the task-specific information of the current scenario. Similarly, this structure computes the weighted sum of all expert networks through the gating network as output. 
In addition, to avoid the task's interference among different scenarios, the output $C_{i} $ of the $i$-th scenario from former layer will be input into the scenario-specific CGC module.
The input ${T}_{i}^{j}$ of the tower unit for the task-$j$ in the $i$-th scenario is as follows: 
\begin{equation}
    {T}_{i}^{j} = \delta_{i}^{j}(C_{i}) \left[ E_{th}^{i}(C_{i}) || E_{tp}^{ij}(C_{i}) \right]
\end{equation}
where $E_{th}^{i}(C_{i})$ and $E_{tp}^{ij}(C_{i})$ denote the set of task-shared experts and task-specific experts (the task-$j$) in the $i$-th scenario. $\delta_{i}^{j}(C_{i})$ is a gating network to calculate the task-$j$'s weight vector of the $i$-th scenario by linear transformation and a softmax layer:
\begin{equation}
    \delta_{i}^{j}(C_{i}) = softmax(W_{i}^{j}C_{i})
\end{equation}
where $W_{i}^{j} \in R^{(m_i+n_i^j) \times {d}^{'}}$ is a parameter matrix, $m_i$ and $n_i^j$ are the number of $E_{th}^{i}(C_{i})$ and $E_{tp}^{ij}(C_{i})$ respectively, ${d}^{'}$ is the dimension of the $C_{i} $.

The predictions of the task-$j$ under the $i$-th scenario is: 
\begin{equation}
    \hat{y}_{i}^{j} = \tau_i^j({T}_{i}^{j})
\end{equation}
where $\tau_i^j(\cdot)$ denotes the tower unit of the task-$j$ under the $i$-th scenario, which is composed of MLP with activation function.
\subsection{Training Objective}
The final loss function of our proposed HiNet is:
\begin{equation}
    Loss=\sum_{i=1}^{M}\sum_{j=1}^{N_i}\lambda_{i}^{j}\cdot Loss_{i}^{j}(y_{i}^{j},\ \hat{y}_{i}^{j})
\end{equation}
where $N_i$ denotes the number of tasks in the current scenario, $\lambda_{i}^{j}$ is a hyper-parameter that controls the proportion of different losses. Note that according to experience, the hyper-parameter $\lambda_{i}^{j}$ is set to the reciprocal of the current scenario dataset proportion. $Loss_{i}^{j}(\cdot)$ is the cross-entropy loss function.

\section{Experiment}
In this section, extensive offline and online experiments are performed
to evaluate the effectiveness of our proposed HiNet model.
We begin by describing the main experimental setup. 
Then, we compare HiNet with representative multi-scenario and multi-task learning models to verify the effectiveness. 
Ablation studies are carried out in order to verify the efficacy of each key component and hyperparameter.
Finally, we show the correlations among scenarios by visualization of SAN.

\subsection{Experiment Setup}
\subsubsection{Dataset}
We collect industrial user log data from six scenarios ($a$ to $f$) at \textit{Meituan Meishi} platform, among which, scenario $a$ and $b$ have much higher user traffic than scenario $c$ to $f$. 
The experimental data statistics details of each scenario is shown in Table~\ref{tab:tb1}.

\subsubsection{Model Comparison \& Implementation Details}
We compare the proposed HiNet with other SOTA models, 

Multi-task learning model: \textbf{Shared Bottom} \cite{caruana1997multitask}, a neural network model with hard parameter sharing. \textbf{MMoE} \cite{ma2018modeling} uses a flexible gating network to adjust the expert network representation information and eventually uses tower units to fuse all expert network representation information for each task. \textbf{PLE} \cite{tang2020progressive} explicitly divides the expert network into task-shared experts and task-specific experts based on MMoE, which effectively alleviates the negative transfer problem caused by the seesaw phenomenon.

Multi-scenario learning model: \textbf{HMoE} \cite{li2020improving} is adapted from MMoE to model the predicted values of multiple scenarios and optimize the task prediction results for the current scenario. \textbf{STAR} \cite{sheng2021one} constructs a shared and scenario-specific network for the learning of the current scenario by the STAR topology.

Actually, the above baseline models were originally proposed only for solving the problem of multi-task/scenario learning.
In order to achieve a fair experimental comparison, we have made adaptive extension in the implementation. 
Specifically, for multi-task/scenario learning models, we set each task in a scenario as a independent task/scenario. 

\subsubsection{Evaluation Metrics}
We consider CTR and CTCVR tasks for each scenario respectively. The area under ROC curve (AUC) is adopted as the evaluation metric. Note that, for each SOTA model, its hyper-parameter setting are tuned in order to achieve the best performance. In addition, experiments of all models are repeated 10 times respectively, and the \textit{Friedman test} is used to verify the validity of performance of models. 

\subsection{Performance Comparison}

The performance of all models are reported in Table~\ref{tab:tb2}. Results show that our proposed HiNet model outperforms the other baseline models in both CTR and CTCVR task metrics for all scenarios, which demonstrates HiNet's superiority in multi-scenario \& multi-task learning.

From the experiments, we find that both multi-task models and multi-scenario models show a clear seesaw phenomenon. For example, in Scenario $a$, the CTCVR of PLE is higher than that of MMoE, but its CTR is lower than that of MMoE. The simultaneous improvement in both tasks has not been achieved.
Moreover, in terms of multiple scenarios, STAR performs higher than HMoE in CTCVR on scenarios $a$ and $b$, but weaker than HMoE in the same metrics on scenarios such as $c$, $d$, $e$, and $f$. The performance gain of the same metric is not realizable jointly in multiple scenarios.
Compared with these models, HiNet can better capture and leverage the correlations among scenario and tasks with the introduction of modules such as SAN and CGC, which effectively counteracts the phenomenon of negative transfer. 
What's more, HiNet achieves considerable improvement in low traffic scenarios such as $c$, $d$, $e$, and $f$, which demonstrates the effectiveness of HiNet for long-tail scenario learning.

\begin{table*}[ht]
  \caption{The experiments on the number of sub-expert/expert networks on the different information extraction layer.}
  \centering
  \label{tab:tb5}
  \resizebox{0.9\linewidth}{!}{
  \begin{tabular}{c|c|cccccccccccc}
    \hline \multirow{2}{*} { Layer } & \multirow{2}{*} { Model } & \multicolumn{2}{c} { Scenario-$a$ } & \multicolumn{2}{c} { Scenario-$b$ } & \multicolumn{2}{c} { Scenario-$c$ }& \multicolumn{2}{c} { Scenario-$d$ }& \multicolumn{2}{c} { Scenario-$e$ }& \multicolumn{2}{c} { Scenario-$f$ } \\
			\cline { 3 - 14 } & & auc\_ctr & auc\_ctcvr & auc\_ctr & auc\_ctcvr & auc\_ctr & auc\_ctcvr & auc\_ctr & auc\_ctcvr & auc\_ctr & auc\_ctcvr & auc\_ctr & auc\_ctcvr \\
    \hline 
    \multirow{4}{*}{Scenario} & HiNet(1 subexpert) & 0.7670 & 0.8485 & 0.7648 & 0.7859 & 0.6699 & 0.6986 & 0.6386 & 0.6766 & 0.5792 & 0.5599 & 0.7644 & 0.7863 \\
    & HiNet(3 subexperts)  & 0.7872 & 0.8668 & 0.7923 & 0.8118 & 0.6983 & 0.7411 & 0.6781 & 0.7216 & 0.6223 & 0.6128 & 0.8013 & 0.8243 \\
    & HiNet(5 subexperts, \textbf{ours}) & \textbf{0.7904} & 0.8675 & \textbf{0.7996} & 0.8194 & 0.7146 & 0.7620 & 0.6858 & 0.7282 & \textbf{0.6253} & 0.6203 & 0.8115 & \textbf{0.8359} \\
    & HiNet(7 subexperts) & 0.7843 & \textbf{0.8677} & 0.7992 & \textbf{0.8206} & \textbf{0.7192} & \textbf{0.7624} & \textbf{0.6887} & \textbf{0.7291} & \textbf{0.6253} & \textbf{0.6214} & \textbf{0.8124} & 0.8354 \\
    \hline
    \multirow{5}{*}{Task} & HiNet(1 expert) & 0.7809 & 0.8607 & 0.7826 & 0.8025 & 0.6879 & 0.7216 & 0.6624 & 0.6967 & 0.5916 & 0.6107 & 0.7882 & 0.81098 \\
    & HiNet(2 experts, \textbf{ours}) & 0.7904 & 0.8675 & 0.7996 & 0.8194 & 0.7146 & 0.7620 & 0.6858 & 0.7282 & 0.6253 & 0.6203 & 0.8115 & 0.8359 \\
    & HiNet(3 experts) & 0.7938 & 0.8705 & 0.8024 & 0.8227 & 0.7142 & 0.7691 & 0.6896 & 0.7288 & 0.6284 & 0.6210 & 0.8180 & 0.8354 \\
    & HiNet(4 experts) & 0.7939 & 0.8746 & 0.8057 & 0.8244 & 0.7205 & 0.7757 & 0.6939 & 0.7301 & 0.6281 & 0.6246 & 0.8202 & 0.8368 \\
    & HiNet(5 experts) & \textbf{0.7950} & \textbf{0.8757} & \textbf{0.8069} & \textbf{0.8266} & \textbf{0.7219} & \textbf{0.7794} & \textbf{0.6956} & \textbf{0.7315} & \textbf{0.6290} & \textbf{0.6243} & \textbf{0.8216} & \textbf{0.8381} \\
    \hline
  \end{tabular}
  }
\end{table*}

\subsection{Ablation Study}

\subsubsection{Key component of HiNet}
To investigate the efficacy of each key component of HiNet, we design five variants for ablation analysis. Specifically, 
(1) \textit{HiNet w/o hierarchy} removes the hierarchical architecture of information extraction layers, and directly adopts CGC to multi-scenario and multi-task learning, 
(2) \textit{HiNet w/o SAN} represents the model removing the SAN component in Scenario Extraction Layer,
(3) \textit{HiNet w/o task gating} denotes the variant that removes the gating network from task information extraction layer,
(4) \textit{HiNet w/o scenario gating} stands for the variant that removes the gating network from scenario information extraction layer,
(5) \textit{HiNet w/o scenario \& task gating} indicates the variant that removes the gating network from both scenario information extraction layer and task information extraction layer.

From Table~\ref{tab:tb4} we can observe that the variant model without hierarchy has serious performance degradation in all metrics, which indicates that hierarchical information extraction structure can effectively capture the commonalities and distinctions across scenarios thus improve the performance of the model.
Analogously, when the SAN module is removed, the performance decreases significantly in multiple scenarios, which suggests that the information derived from the SAN module can effectively complement the representation learning of scenarios.
In addition, variants-(3), (4), and (5) of the model validate the effect of the gating network on the representational ability of the model, which shows that the gating network can flexibly adapt
the weights of expert networks, which enables HiNet to effectively distinguish important information
from representations of expert networks.

\subsubsection{The hyperparameters of sub-expert/expert network}
Hyperparameters such as the number of expert networks may affect the performance of HiNet. So we set up experiments with different hyperparameters as follows: (1) the experiment on different numbers of sub-expert networks, (2) the experiment on different numbers of expert networks. Note that in our final model, we set the number of sub-expert networks of the SEI module to be 5, and the number of shared and task-specific expert networks to be 2.

From Table \ref{tab:tb5}, it can be seen that the model performance improves as the number of sub-expert networks or expert networks increases. 
Intuitively, the number of parameters becomes larger when HiNet is equipped with more expert networks.
However, a larger model normally requires more training data. 
Therefore, the number of expert networks can be regarded as a hyper-parameter to control the trade-off between the practical performance and model size.

\subsection{Model Complexity}
Since scalability is very important for industrial applications, we compared different models' parameters, training time, and inference time (in all scenarios). 
All our experiments are performed on a workstation equipped with an NVIDIA Tesla A100 GPU, 80G RAM, and Intel(R) Xeon(R) Gold 5218 CPU. 
Table \ref{tab:tb7} compares the results. 
From Table \ref{tab:tb7}, for HiNet and its competitors, the differences are not significant regarding the numbers of parameters, the training time, and the inference time, suggesting that our HiNet model is also applicable in practice and does not bring the additional cost in real-world applications.

\begin{table}[ht]
  \caption{Time and space complexity of HiNet and its competitors on the Meituan dataset.
  }
  \centering
  \label{tab:tb7}
  \resizebox{1.0\linewidth}{!}{
  \begin{tabular}{c|c|c|c}
    \hline Model & Params($\times 10^7$) & Training time (minutes) & Inference time (milliseconds/sample) \\
    \hline 
    MMoE & 1.94 & 163.29 & 0.1447 \\
    PLE & 2.05 & 175.05 & 0.1688 \\
    HMoE & 2.03 & 167.88 & 0.1752 \\
    STAR & 2.11 & 177.62 & 0.1825 \\
    HiNet & 2.07 & 171.07 & 0.1731\ \\
    \hline
  \end{tabular}
  }
\end{table}

\subsection{Visualization of Scenario-aware Attentive Network}

\begin{figure}[t]
  \centering
  \includegraphics[width=0.9\columnwidth]{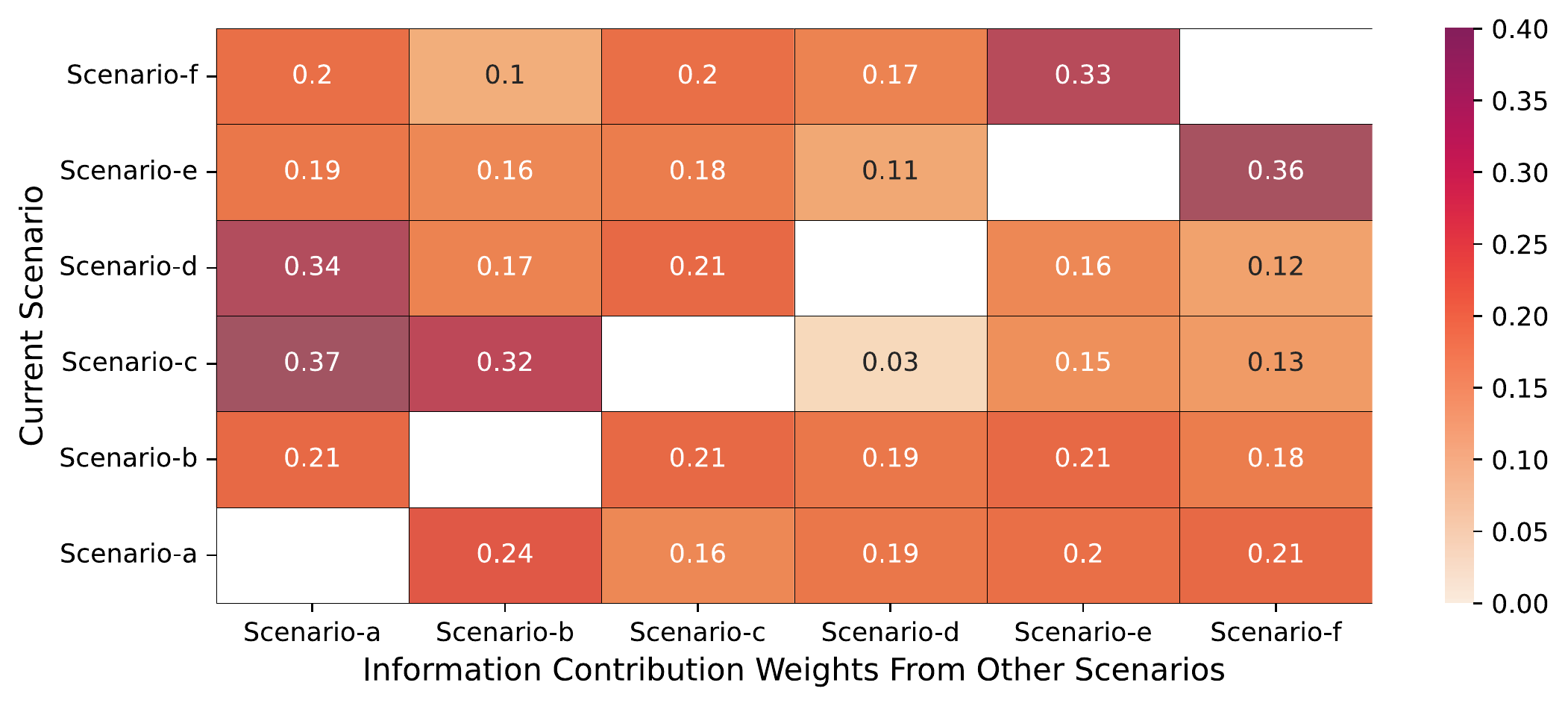}
  \caption{Visualization of correlation weights among different scenarios.}
  \label{fig:heatmap}
\end{figure}

In order to intuitively demonstrate the effectiveness of the SAN module, we conduct a visual analysis to explicitly explore the varying relevance between scenarios.
Specifically, we display the final information importance scores output by the softmax layer of the SAN module on a heat map (as shown in Fig.~\ref{fig:heatmap}), where darker colors represent higher importance.
This figure obviously reveals that, scenarios have varying degrees of contribution to each other, which suggests a measure of correlation among different scenarios.
\section{Online A/B Test}
To further validate the online performance of the proposed model, we deployed HiNet in \textit{Scenario-$a$} and \textit{Scenario-$b$} at Meituan, and conducted a one-month online A/B test against baseline model. 
The results of online A/B test show that: a) in Scenario-$a$, compared with the baseline model, HiNet achieved relative improvement of 0.24\% and 3.38\% in CTR and CTCVR; 
b) in Scenario-$b$, HiNet obtained relative improvement of 0.11\% and 1.47\% in CTR and CTCVR.
The online results clearly demonstrate that our proposed HiNet model can bring considerable practical benefit. Currently, HiNet has served as the major traffic model for both Scenario-$a$ and Scenario-$b$, which makes a significant improvement by +2.87\% and +1.75\% in terms of order quantity.

\section{Conclusion}
Multi-scenario \& multi-task modeling is one of the most crucial and challenging problems in recommendation systems. Previous models mainly focus on optimizing multiple tasks in different scenarios by projecting all information into the same feature space, which leads to unsatisfactory performance. In this paper we conduct in-depth analysis based on industrial data from Meituan Meishi platform and propose HiNet, which utilizes hierarchical optimization architecture to model the multi-scenario \& multi-task problem. On this basis, the SAN module is designed to enhance the ability of representation learning of scenarios. Both offline and online experiments validate the superiority of the HiNet model. Currently, HiNet has been fully deployed in recommendation systems at Meituan for online serving.

\section*{Acknowledgements}
This work was supported by the National Natural Science Foundation of China (No. 62202025 and 62002012).

\balance
\bibliographystyle{IEEEtran}
\bibliography{IEEEabrv,reference}

\end{document}